\documentclass[10pt,letterpaper]{article}

\usepackage{keyval}
\usepackage{opex3}
\usepackage{geometry}
\usepackage{subfig}
\usepackage{caption}
\DeclareCaptionLabelSeparator{dot}{.~}

\newcommand {\micro}[1]{$\mathrm{\mu}$#1}

\newcommand {\corresponding} [1]{\vskip-.3cm \parskip0pc\hskip2.25pc \footnotesize%
   \parbox{.8\textwidth}{\begin{center} $^{\ast}$Corresponding author: \it \textcolor{blue}{\underline{#1}} \rm \end{center} } \normalsize  \vskip-.2cm}

\begin{document}

\title{Testing ultrafast mode-locking at microhertz relative
optical linewidth}

\author{Michael J. Martin$^{1, \ast}$, Seth M.
Foreman$^{1,2}$, T. R. Schibli$^{1}$, and Jun Ye$^{1}$}

\address{$^{1}$JILA, National Institute of Standards and Technology and University of Colorado\\
               Department of Physics, University of Colorado, Boulder, CO 80309--0440\\
        $^{2}$Current address: Varian Physics, Room 246, 382 Via Pueblo Mall, Stanford, CA 94305-4060

}

\corresponding{Michael.J.Martin@colorado.edu}



\begin{abstract}
We report new limits on the phase coherence of the ultrafast
mode-locking process in an octave-spanning Ti:sapphire comb. We
find that the mode-locking mechanism correlates optical phase
across a full optical octave with less than 2.5 \micro Hz
relative linewidth. This result is at least two orders of
magnitude below recent predictions for quantum-limited
individual comb-mode linewidths, verifying that the
mode-locking mechanism strongly correlates quantum noise across
the comb spectrum.
\end{abstract}



\section{Introduction}
Optical frequency combs have become ubiquitous tools for
precision optical measurement~\cite{udem2002,
RevModPhys.75.325}. They have enabled a new generation of
optical frequency references based on narrow transitions in
single trapped ions~\cite{oskay2006sao, margolis2004hlm,
PhysRevA.70.012507, rosenband2008fra} and cold, neutral, atomic
ensembles~\cite{ludlow2008slc, poli2008fed, katori2005nature,
sterrNistPTB, ludlow2006sss, letargat2006aol, campbell-2008}.
In addition, the comb's ability to phase coherently transfer
optical references across large spectral gaps allows for direct
optical atomic clock comparison, placing new constraints on the
evolution of fundamental constants~\cite{rosenband2008fra} as
well as driving atomic clock technology~\cite{ludlow2008slc}.
Recently, this broadband phase coherence has facilitated the
production of ground-state ultracold polar molecules via
stimulated Raman adiabatic passage \cite{K.-K.Ni10102008,
JohannG.Danzl08222008}.

Previous evaluations have proven the frequency comb's
suitability for precision optical metrology. The fractional
frequency uncertainty of Ti:sapphire-based frequency combs has
been evaluated at the $10^{-19}$ level at
1000~s~\cite{ma2004ofs, Ma2007}. Experiments testing the phase
coherence of Ti:sapphire combs have been able to place upper
limits on the relative linewidth of different spectral regions
for both locked and free running combs at 20
mHz~\cite{bartels2004sfl} and 9~mHz~\cite{stenger2002umo},
respectively, and were ultimately limited by by
differential-path technical noise caused by air currents and
mirror vibrations. The frequency comb used in this work was
previously compared to a 10~W average power Yb fiber comb
locked to a common optical reference, with a resulting 1~mHz
resolution bandwidth-limited relative
linewidth~\cite{schibli2008ofc}. This indicates that the
Ti:Sapphire comb should be capable of supporting narrower
relative linewidths. Here we report a new lower limit to the
intrinsic phase coherence of a mode-locked Ti:sapphire laser
phase locked to an optical reference, by both linewidth and
phase noise measurements. This work demonstrates that the
mode-locking process correlates the phase noise of individual
frequency comb modes at a level far below the quantum noise
limit of individual free-running comb modes.

A complete description of the frequency of a given output mode
of a comb is given by
\begin{equation}
\nu_{n}(t) = n f_{\mathrm{rep}}(t) + f_{0}(t) + \epsilon_{n}(t).
\label{Basic_comb}
\end{equation}
Here $n$ is the index labeling the harmonic of the repetition
rate $f_{\mathrm{rep}}(t)$ and $f_{0}(t)$ is the carrier
envelope offset frequency. Both $f_{\mathrm{rep}}(t)$ and
$f_{0}(t)$ are radio frequency (RF) signals that can be
measured by directly observing the output pulse on a
photodetector, and by employing a self-referencing technique
(\textit{e.g.}~\cite{jones2000cep}), respectively. The term
$\epsilon_{n}(t)$ represents frequency noise in the vicinity of
mode $n$ that is not described by fluctuations in
$f_{\mathrm{rep}}(t)$ or $f_{0}(t)$. In other words,
$\epsilon_{n}(t)$ accounts for mode-dependent noise terms that
are at least quadratic in order with respect to $n$ as a result
of pulse-to-pulse fluctuations, which could be in part caused
by time-dependent fluctuations of higher-order intracavity
dispersion. Additionally, and most importantly,
$\epsilon_{n}(t)$ is also assumed to include fluctuations not
related to any other comb mode---fluctuations that are
completely uncorrelated across the comb. In this way,
$\epsilon_{n}(t)$ also accounts for possible spontaneous
emission-induced frequency noise that does not affect the
global timing and phase parameters of the comb as a result of
imperfect mode-locking. This is in contrast to the case of an
ideal comb, where the mode $n$ is perfectly defined with
respect to mode $m$ in the limit where both $\epsilon_{m}$ and
$\epsilon_{n}$ are zero, due to the mode locking mechanism
enforcing a fixed phase relation across the comb. While the
relative coherence of comb teeth is only limited by the quality
of the mode-locking process, a free-running frequency comb,
when compared to an external reference, has noise properties
dominated by vibrational noise in the mirror mounts and thermal
drifts in the laser cavity coupling to both $f_{\mathrm{rep}}$
and $f_{0}$.


Actively phase locking a Ti:sapphire comb to an optical
reference requires control of both $f_{\mathrm{rep}}$ and
$f_{0}$ via control of the laser cavity length and pulse group
delay (\textit{e.g.}~\cite{reichert1999mfl, ye2000ppc}).
Combining the comb and a continuous wave (CW) laser results in
a time-dependent heterodyne beat, formed when a given comb
tooth $n$ interferes with the CW laser. This RF signal is
denoted by $f_{b,n}(t)$. By phase locking $f_{b,n}(t)$ to an RF
source, the comb ideally acquires the optical phase information
of the reference laser and the RF source. This is due to the
fixed phase relationship between the comb's output modes,
enforced by the mode-locking process. When locked via control
of $f_{\mathrm{rep}}$, $f_{b,n}(t)$ is related to the RF
reference frequency, $f_{\mathrm{RF}}$, and the comb degrees of
freedom by
\begin{equation}
f_{b,n}(t) = f_{\mathrm{RF}}+ \delta f_{b,n}(t) = nf_{\mathrm{rep}}(t)+f_{0}(t) + \epsilon_{n}(t) -\nu_{\mathrm{CW}}(t). \label{N_beat}
\end{equation}
Here, $\delta f_{b,n} (t)$ is the locking error due to the
finite gain of the servo, and $\nu_{\mathrm{CW}}(t)$ is the
frequency of the CW optical reference. We defer consideration
of shot noise, which adds a white phase noise term to the right
side of Eq.~(\ref{N_beat}), to Section 3. Additionally locking
$f_{0}$ to an RF reference, with locking error $\delta
f_{0}(t)$, constrains both comb degrees of freedom. Solving for
$f_{\mathrm{rep}}(t)$ yields

\begin{equation}
f_{\mathrm{rep}}(t) = \frac{1}{n} \left[f_{\mathrm{RF}} + \delta f_{b,n}
(t) - \epsilon_{n}(t) +\nu_{\mathrm{CW}}(t) - f_{0} -\delta f_{0}(t)  \right]. \label{frep}
\label{frep}
\end{equation}
It is important to note that the locking errors and noise term
$\epsilon_{n}(t)$ write noise onto $f_{\mathrm{rep}}(t)$, thus
globally affecting the comb. Using Eq.~(\ref{Basic_comb}) and
$f_{\mathrm{rep}}(t)$ given in Eq.~(\ref{frep}), the optical
frequency of a comb mode numbered $m$ is given by
\begin{equation}
\nu_{m} = \frac{m}{n} \left[f_{\mathrm{RF}} + \delta f_{b,n}
(t) - \epsilon_{n}(t) +\nu_{\mathrm{CW}}(t) \right] + \left(1-\frac{m}{n}\right)\left[f_{0} +\delta f_{0}(t)\right]+ \epsilon_{m}(t).
\label{nu_m}
\end{equation}
Again the added noise term, $\epsilon_{m}(t)$, represents extra
frequency noise added to mode $m$ by both correlated and
uncorrelated laser dynamics. Thus, measuring the comb mode $m$
relative to mode $n$ in a precise way allows upper limits on
the intrinsic noise properties of the comb to be determined.
One way to accomplish this, as we report in this work, is to
use the second harmonic of the optical reference to which mode
$n$ is locked to compare the phase coherence of comb modes $n$
and $m=2n$ in a direct way. In this case, the expected
heterodyne beat between the second harmonic light and the comb
mode $2n$, which represents an out-of-loop measurement of the
relative coherence of modes $n$ and $2n$, is given by
\begin{equation} f_{b,2n}(t) = \nu_{2n}(t) - 2\nu_{\mathrm{CW}}(t) = 2 f_{\mathrm{RF}} - f_{0}  - \delta
f_{0}(t) + 2 \delta f_{b,n}(t) + \epsilon_{2n}(t) - 2
\epsilon_{n}(t). \label{OL_beat}
\end{equation}
Here, the term $\epsilon_{2n}(t) - 2 \epsilon_{n}(t)$
represents the time-fluctuating out-of-loop frequency noise
added by the comb dynamics.


The Wiener--Khinchin theorem relates time-domain frequency
fluctuations to single-sided frequency noise power spectral
density by
\begin{equation}
S_{\nu}(f) = 4 \int^{\infty}_{0} \cos\left(2 \pi \tau f\right) R_{\xi \xi}(\tau) d \tau .
\label{WK}
\end{equation}
Here, $S_{\nu}(f)$ is the power spectral density associated
with the time-fluctuating frequency $\xi(t)$. The
autocorrelation term $R_{\xi \xi}(\tau)$ is defined by
\begin{equation}
R_{\xi \xi}(\tau) = \langle \xi(t) \xi(t+\tau) \rangle = \lim_{T \to \infty} \frac{1}{T}\int^{T/2}_{-T/2}\xi(t) \xi(t+\tau) dt.
\label{R}
\end{equation}
Additionally, $S_{\nu}(f)$ is related to phase power spectral
density by
\begin{equation}
S_{\phi}(f)=S_{\nu}(f)/f^{2}.
\label{ftophi}
\end{equation}

In the case that $\xi(t)$ represents the noise
$\epsilon_{2n}(t) - 2 \epsilon_{n}(t)$, then Eq.~(\ref{WK})
provides a description of the frequency noise power spectral
density induced by this term on the out-of-loop beat.
Equation~(\ref{WK}) will include the possible effects of
correlated dynamics between $\epsilon_{2n}(t)$  and $
\epsilon_{n}(t)$, which add in quadrature with the completely
uncorrelated components of $\epsilon_{2n}(t)$  and $
\epsilon_{n}(t)$. Thus, $S_{\nu}(f)$ of Eq.~(\ref{WK}), with
$\xi(t) =\epsilon_{2n}(t) - 2 \epsilon_{n}(t)$, represents an
upper limit to the completely uncorrelated noise between modes
$n$ and $2n$.


%


\section{Experiment}

The basic approach we take for our measurement of the
out-of-loop coherence between modes $n$ and $2n$ is to
stabilize the comb to a CW laser, and then use the second
harmonic of the same CW laser as a reference. When compared
against the comb, this second harmonic reference enables the
phase coherence of the comb across a full optical octave to be
tested. The optical phase lock to the CW laser is implemented
by servo control of the laser cavity length and pump power,
while an $f$--$2f$ interferometer provides the additional
signal used to stabilize $f_{0}$. The heterodyne beat between
the comb and second harmonic CW reference represents the
out-of-loop signal, which includes contributions from both the
$f_{\mathrm{rep}}$ and $f_{0}$ phase locked loops as described
by Eq.~(\ref{OL_beat}). Any differential-path effects limit the
sensitivity of this measurement, and we have taken care to
limit their effect by careful design.

\begin{figure}
\center
\includegraphics[scale = .5]{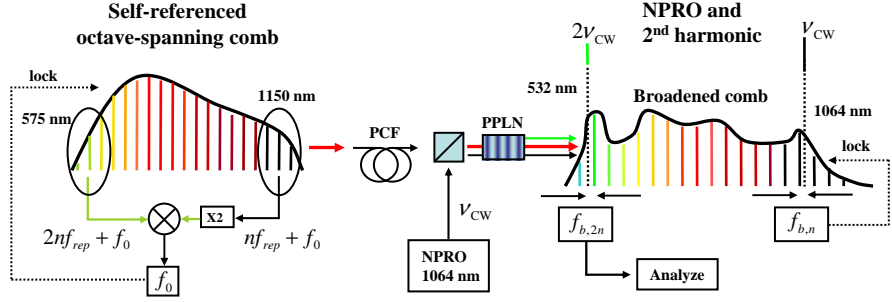}
\captionsetup{margin=0.5in,font=footnotesize, labelsep=period}
\caption{After passing through an $f-2f$ interferometer for self-referencing, the remaining spectrum (600--1100~nm) from the octave-spanning frequency comb is
broadened in photonic crystal fiber (PCF) and overlapped with
the output of an NPRO Nd:YAG at 1064~nm. The resulting heterodyne beat, $f_{b,n}$, and the carrier envelope offset frequency, $f_{0}$, are locked to RF references. Polarization-selective
doubling of the overlapped beams ensures that only the CW light is
frequency doubled, while collinear beam propagation reduces
technical noise. The out-of-loop  beat at 532~nm, $f_{b,2n}$, is
analyzed as detailed in the text.} \label{Fig:Experiment}
\end{figure}

The specific octave spanning Ti:sapphire frequency comb used in
this experiment is similar to the system described in
\cite{fortier2003pso}. The relatively low repetition rate of
95~MHz leads to high pulse energy. This facilitates self phase
modulation in the laser crystal, which broadens the spectrum to
a full octave, the wings of which are not resonant with the
cavity and are immediately transmitted by the output coupler.
As detailed in Fig.~\ref{Fig:Experiment}, approximately 3~nm
wide spectral regions at 575~nm and 1150~nm are used to measure
$f_{0}$ with a standard $f$--$2f$ interferometer. This RF
signal is used to lock $f_{0}$, via group delay actuation
(using the same method as described in \cite{reichert1999mfl}),
to an RF source that shares a common timebase with all the RF
sources in the experiment. The remaining optical spectrum that
is not used to measure $f_{0}$ consists of 600--1100~nm light,
which is rebroadened to an optical octave centered near 750~nm
using photonic crystal fiber (PCF). The output of the PCF is
combined with a Nd:YAG non-planar ring oscillator (NPRO) CW
optical reference at 1064~nm with polarization orthogonal to
the comb. After passing through optical band-reject filters to
remove the majority of the comb power in the unused central
portion of the comb spectrum, the co-propagating comb and
500~mW of 1064~nm light pass through a temperature stabilized
periodically poled lithium niobate (PPLN) crystal, doubling the
1064~nm CW light while overlapped with the comb, reducing
technical noise due to differential path effects. The 1064~nm
comb light is not doubled due to its orthogonal polarization,
although if it were the resulting RF beat would be
distinguishable from the true out-of-loop signal by its central
frequency. A $\lambda/2$ at 1064~nm, $\lambda$ at 532~nm wave
plate is placed before the PPLN crystal in order to ensure that
the the second harmonic CW light has the same polarization as
the 532~nm comb light, since the PPLN outputs parallel
polarizations of second harmonic and fundamental CW light.  A
single beam exits the PPLN crystal, with the 1064-nm comb light
polarized orthogonally to the other three components of
interest.


A Glan-Thompson polarizer separates the majority of the 1064~nm
comb light and $\sim$1~mW of the CW 1064~nm light, allowing
measurement of the RF heteredyne beat at 1064~nm. As described
by Eqs.~(\ref{N_beat}) and (\ref{frep}), this signal is used to
stabilize the comb mode at $\nu_n$ ($n \simeq 3 \times 10^{6}$)
by phase locking the heterodyne beat, via cavity length and
pump power control, to an RF synthesizer. The remaining CW
light at 1064 is transmitted by the polarizer, along with the
majority of the comb and CW component at 532~nm.

The transmitted light contains approximately 500~\micro W of
second harmonic CW light and comb near 532~nm, which is
filtered to reject a large component of residual fundamental CW
light. We measure the resulting heterodyne beat between the
comb mode $2n$ and the second harmonic CW light. The expected
RF frequency is given by Eq.~(\ref{OL_beat}), where
$\nu_{\mathrm{CW}}$ is now specifically referring to the
frequency of the NPRO. In principle, the time-domain frequency
error due to finite servo gain, given by
\begin{equation}
\delta f_{\mathrm{lock}}(t)=-\delta f_{0}(t) + 2 \delta f_{b,n}(t),
\label{Lock_error}
\end{equation}
can be estimated from the in-loop phase error signals and is
indistinguishable from fundamental noise described by the term
$\epsilon_{2n}(t) - 2 \epsilon_{n}(t)$. The phase noise power
spectral density associated with this term can be expressed via
Eqs.~(\ref{WK}--\ref{ftophi}) as
\begin{equation}
S_{\phi}(f) = \frac{4}{f^{2}} \int^{\infty}_{0} \cos\left(2 \pi f \tau\right) \left[R_{\delta f_{0}\delta f_{0}}(\tau) +4R_{\delta f_{b,n}\delta f_{b,n}}(\tau) - 4R_{\delta f_{0}\delta f_{b,n}}(\tau)\right]  d \tau .
\label{Lock_error2}
\end{equation}
If the cross-correlation term, $R_{\delta f_{0}\delta
f_{b,n}}(\tau)$, is zero,  the expected contribution of the
locking error to the out-of-loop noise can thus be estimated by
a weighted sum of the in-loop phase noise power spectral
densities. This will only occur if there are no common noise
sources for the servos locking $f_{0}$ and the comb to the NPRO
reference.

\begin{figure}
\center
\subfloat{\label{Full_span}\includegraphics[scale=.5]{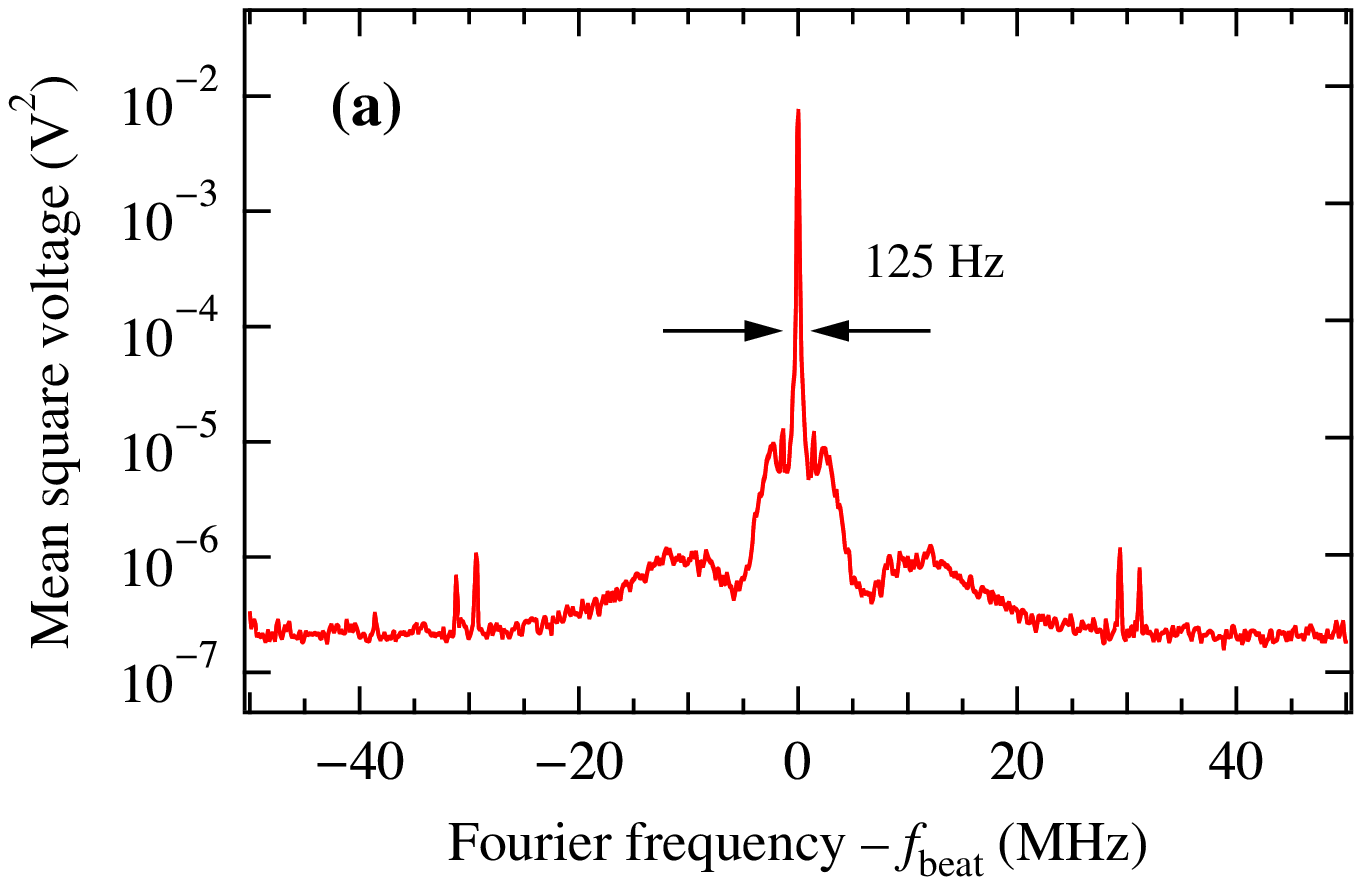}}
\subfloat{\label{FFT_Limit}\includegraphics[scale=.5]{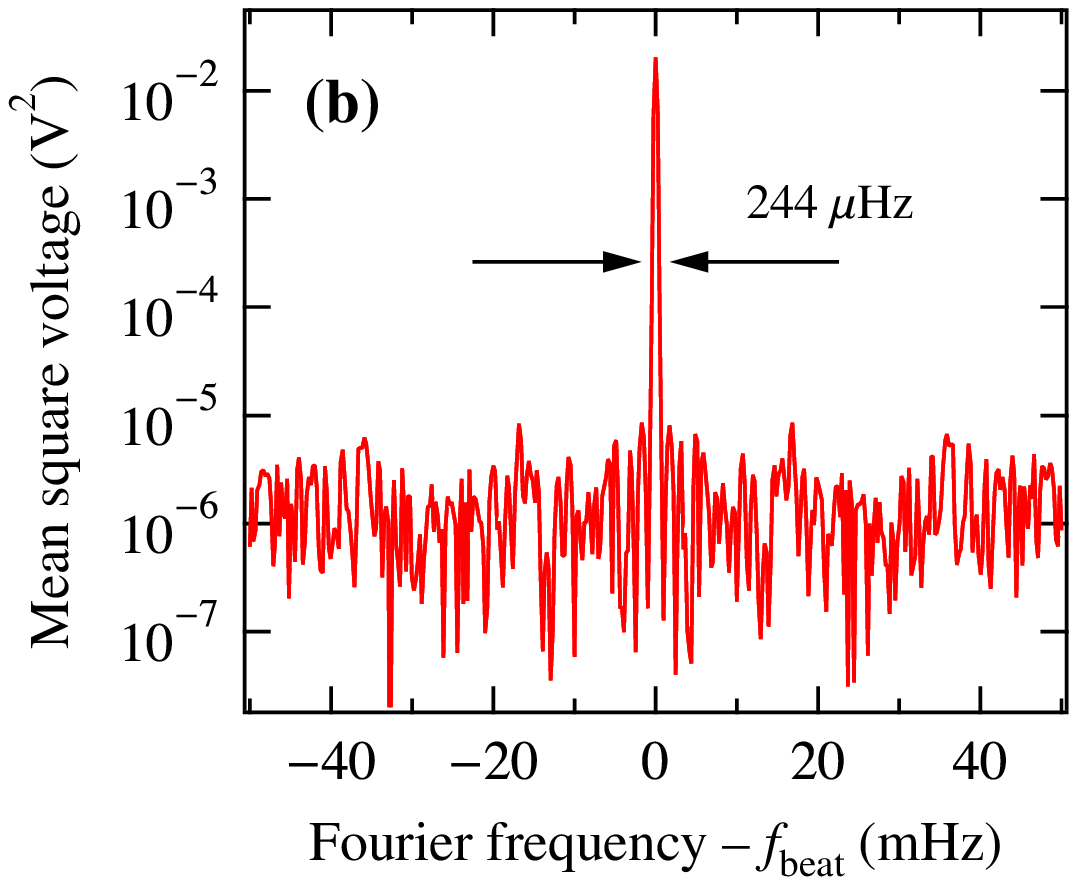}}\\
\captionsetup{margin=0.5in,font=footnotesize, labelsep=period}
\caption{\textbf{(a)} Out-of-loop beat mixed to
50~kHz at 100~kHz span with 125~Hz resolution bandwidth.
Servo bumps are evident due to the contribution of finite
locking gain and bandwidth to $\delta f_{\mathrm{lock}}(t)$.
\textbf{(b)} Out-of-loop beat mixed to 1~Hz at 100~mHz span and
244~\micro Hz resolution bandwidth. A strong coherent carrier is evident, even
in this narrow resolution bandwidth.} \label{FFTs}
\end{figure}

While Eq.~(\ref{Lock_error2}) gives a prediction for the
out-of-loop frequency noise from the in-loop locking error,
there are two additional important sources of out-of-loop noise
that are indistinguishable from fundamental comb noise. Shot
noise propagates through the optical phase locks for both
$f_{b,n}$ and $f_{0}$, and adds in quadrature with the shot
noise on the detector for $f_{b,2n}$, as discussed in
Section~3. Out-of-loop technical noise such as differential
path Doppler noise or amplitude to phase conversion in the
PCF~\cite{fortier2002npn} also contributes to the measured
out-of-loop phase noise spectrum.

One important feature of Eq.~(\ref{OL_beat}) is that it does
not depend on the frequency of the NPRO, which, despite being
quite stable due to its monolithic construction, has drifts on
the order of 1~MHz/min. A shift of 1~MHz will show up on the
1~Hz level in $f_{b,m}(t)$ if $m=2n + 1$, \textit{i.e.} the
conjugate beat is used. This is clearly unacceptable for
measuring comb linewidth on the \micro Hz level. Thus, only
$f_{b,2n}(t)$ is considered.

Figure~\ref{FFTs} shows fast Fourier transforms (FFTs) of
$f_{b,2n}(t)$, after it has been mixed down to near DC. Figure
\ref{FFTs}(b) represents the narrowest resolution bandwidth
obtainable by the FFT instrument used in this experiment, due
to an instrument-limited measurement time of $\sim1$~hr.
Obtaining better resolution bandwidth could be achieved by
digitally sampling and recording the data over a longer time
period and subsequently performing the FFT. However, at time
scales over one hour, the microstructure fiber alignment
drifts, causing the signal to noise ratio  of the out-of-loop
beat to drop well below 20~dB in a 100~kHz bandwidth. Servo
unlocks also occur on this timescale due to finite servo range
and thermal drift.

\begin{figure}
\center
\includegraphics[scale = .5]{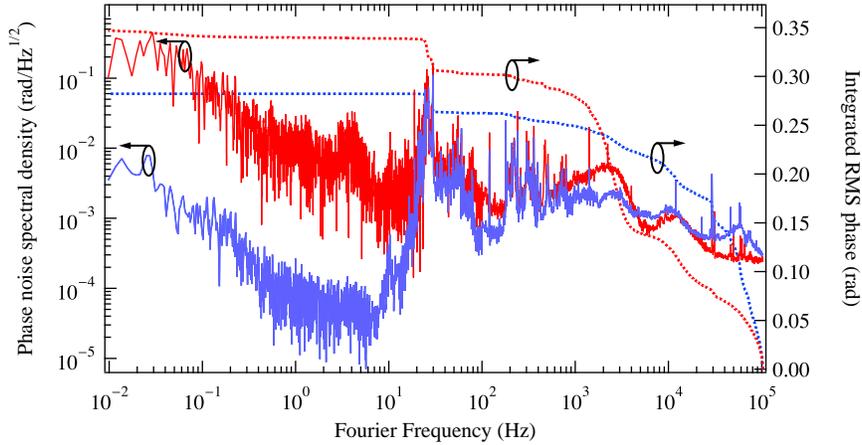}
\captionsetup{margin=0.5in,font=footnotesize, labelsep=period}
\caption{Left axis: Estimate of the out-of-loop phase noise contribution
of $\delta f_{\mathrm{lock}}$ based on a weighted incoherent sum of
in-loop phase noise power spectral densities of both servo
loops (blue) compared to the measured overall out-of-loop phase noise
spectral density (red). Right axis: Root mean square (RMS)
phase error integrated down from 100~kHz for both the
contribution of the servo errors on out-of-loop noise (dashed blue) and
the measured out-of-loop phase noise spectral density (dashed red).}
\label{Phase_noise}
\end{figure}

The measured single-sided phase noise spectral density of the
out-of-loop beat, shown in Fig.~\ref{Phase_noise}, is
complimentary to the linewidth measurement. The out-of-loop
phase noise is directly visible in the sidebands of
Fig.~\ref{FFTs}(a). We note that the integrated
root-mean-square (RMS) phase is 0.35~rad when integrated down
from 100~kHz to 10~mHz. This result, when combined with
Fig.~\ref{FFTs}(b)---which shows no significant features 50~mHz
away from the carrier---indicates that the 244~\micro Hz
instrument-limited coherent linewidth in Fig. \ref{FFTs}(b) is
a robust upper limit to the beat linewidth. The estimated
locking error contribution to the phase noise from typical
error signal spectral densities is additionally shown in
Fig.~\ref{Phase_noise}. Here, Eq.~(\ref{Lock_error2}) has been
used with the assumption that $R_{\delta f_{0}\delta
f_{b,n}}(\tau) \to 0$.
The discrepancy between the predicted out-of-loop phase noise
near 50~kHz is due to the servo bump of the RF tracking filter
used in the $f_{0}$ phase lock. The $f_{0}$ servo does not have
the bandwidth to track this noise, so it appears only on the
in-loop spectrum. Further discrepancies at Fourier frequencies
in the 1~kHz range show the inadequacy of the assumption that
$R_{\delta f_{0}\delta f_{b,n}}(\tau) \to 0$, reflecting the
fact that the cavity length and group delay servos are coupled
when used to lock the comb to an optical reference and
additionally may share common technical noise sources. When
integrated from 100~kHz to 10~mHz, the expected integrated RMS
phase predicted by the in-loop locking error is 70~mrad below
the observed out-of-loop integrated phase error.

To overcome the measurement time-limited resolution bandwidth
of 244 \micro Hz and examine the out-of-loop beat at
significantly improved phase noise sensitivity, we take the
approach of studying the phase noise of a harmonic of the
out-of-loop beat. A step recovery diode impedance matched at
100~MHz generates over 30 harmonics, amplifying the phase
noise. The diode generates a $\sim 100$~ps pulse for every
high-to-low voltage zero crossing, and this output can be
represented in the time domain as

\begin{equation}
V(t) = V_{0}\sum_{k=-\infty}^{\infty} \Lambda\left[t-k/f_{\mathrm{in}} - \Delta \phi(t)/2\pi\right] \simeq V_{0} \sum_{n=-\infty}^{\infty} a_{n} \exp\left[i 2\pi n f_{\mathrm{in}} t - i n\Delta \phi(t)\right].
\label{Fourier_series}
\end{equation}
Here, $\Lambda(t)$ is a temporally narrow function compared to
the inverse input frequency, $f_{\mathrm{in}}$, and $a_{k}$ are
the Fourier series coefficient for $\Lambda(t)$. The above
approximation only holds if the phase can be approximated as
stationary on the timescale of the envelope width. In the limit
where $\Lambda(t)$ is a perfect Dirac delta function,
Eq.~(\ref{Fourier_series}) is exact. The final sum in
Eq.~(\ref{Fourier_series}) shows that for harmonic $n$ the
phase noise power spectral density, $S_{\phi}^{n}(f)$, is
related to that of the fundamental, $S_{\phi}(f)$, by
\begin{equation}
S_{\phi}^{n}(f) = n^{2}S_{\phi}(f).
\label{n_squared}
\end{equation}

\begin{figure}
\center
\subfloat{\label{SRD_scheme}\includegraphics[scale=.5]{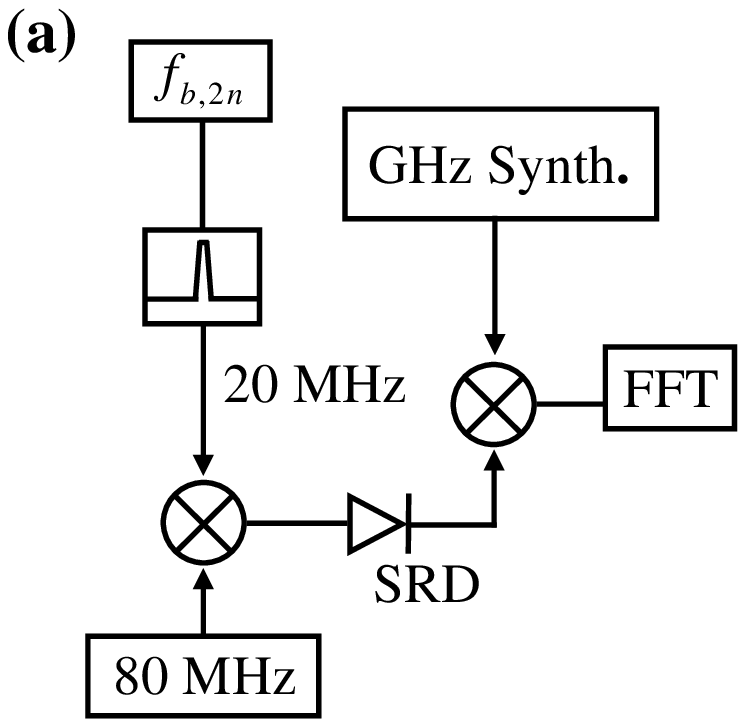}}
$\; \; \; \; \; \; \; \; \; \; \; \;\;\;\;\;\;\; $ \subfloat{\label{Lorentzians}\includegraphics[scale=.5]{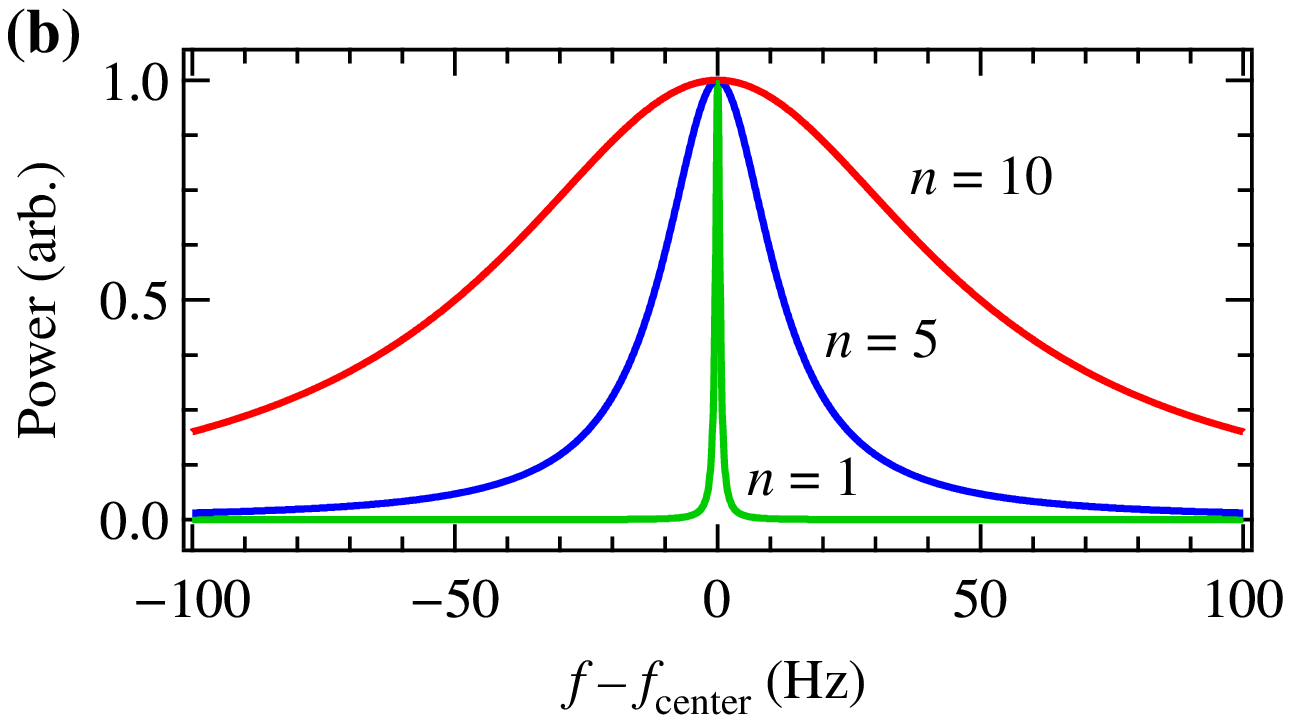}}\\
\subfloat{\label{OLx10}\includegraphics[scale=.5]{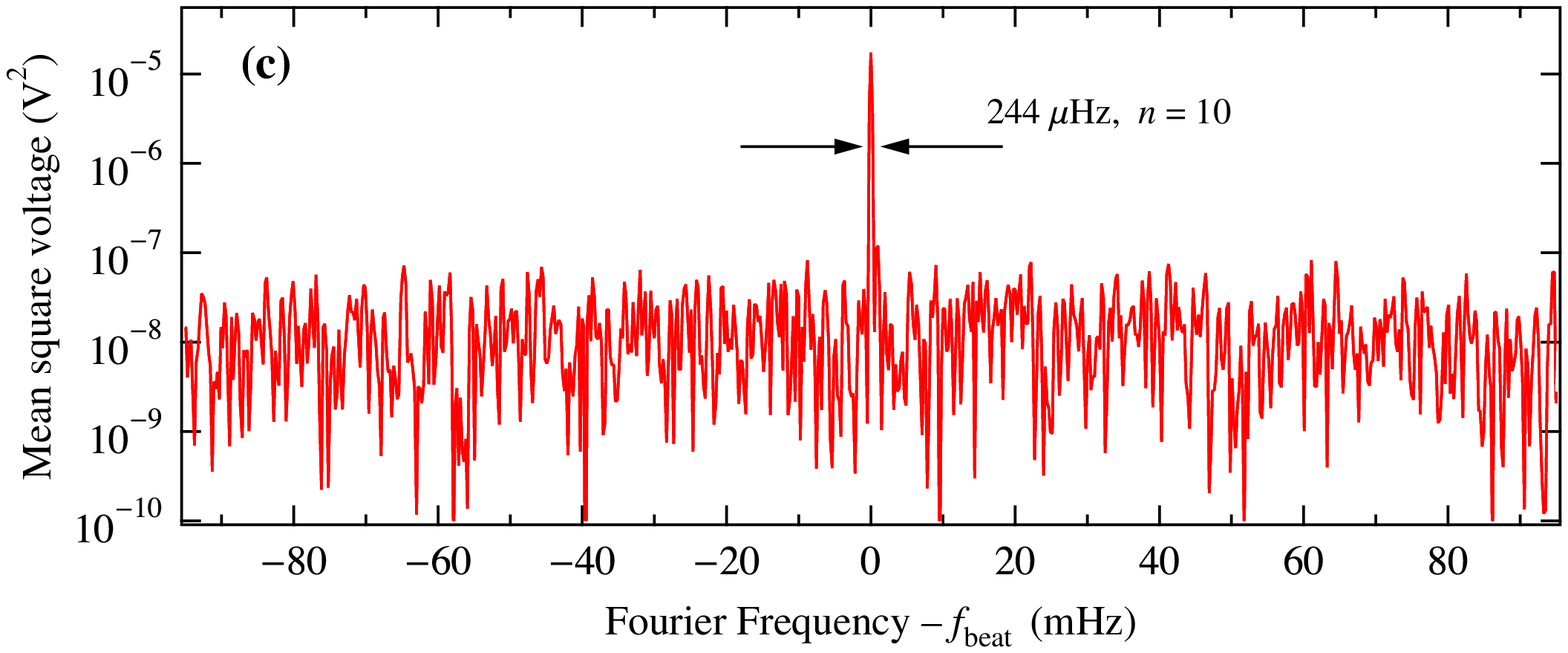}}
\captionsetup{margin=0.5in,font=footnotesize, labelsep=period}
\caption{\textbf{(a)} RF electronics for analyzing the tenth
harmonic of the out-of-loop beat. The 20~MHz signal is filtered with a
2~kHz passband crystal filter and mixed to 100~MHz before being
sent to the step-recovery diode (SRD). The tenth harmonic is
selected for analysis. \textbf{(b)} Schematic depiction of a
Lorentzian lineshape (green) broadened by frequency
multiplication factors of $n=5$ (blue) and $n=10$ (red), yielding linewidths
enhanced by a factor of 25 and 100, respectively. \textbf{(c)}
Out-of-loop signal, 190 mHz span with a factor of 10 frequency
multiplication. This corresponds to a 100-fold increase of the
phase noise power spectral density in the vicinity of the
carrier, resulting in a 100-fold increase in linewidth, yet a
strong coherent peak is still observed in the 244 \micro Hz
resolution bandwidth.} \label{SRD}
\end{figure}

In order to focus on the phase noise nearest the carrier, a
narrow crystal filter centered at 20~MHz with a 2~kHz passband
and 3~dB maximum ripple rejects phase and amplitude noise
greater than 1~kHz away from the carrier. This further enforces
the assumption of stationary phase noise on the time scale of
$1/f_{\mathrm{in}}$ and prevents broadband phase noise from
causing carrier collapse when it is multiplied by the diode.
Using the system whose key components are shown schematically
in Fig. \ref{SRD}(a), we select the 10th harmonic of the step
recovery diode and mix it down to near DC. Figure \ref{SRD}(c)
shows the beat note is still resolution bandwidth limited, even
with the 100-fold increase in phase noise power spectral
density.

\section{Discussion}
From Fig.~\ref{SRD}(c), it is clear that there is no
significant phase noise near the carrier. In order to use this
measurement to extrapolate an upper limit to the actual
non-instrument limited linewidth,  we choose a functional form
of the phase spectral density corresponding to a random walk in
phase,
\begin{equation}
S_{\phi}(f) = \frac{C}{f^{2}}.
\end{equation}
This type of phase diffusion is found due to spontaneous
emission in laser systems, resulting in the famous
Schawlow-Townes limit \cite{schawlow1958iao, goldberg1991tfl}
and would arise if the comb were limited by an ``intrinsic
linewidth'' due to random-walk relative phase diffusion amongst
comb modes. The optical power spectral density corresponding to
this form of phase noise is a Lorentzian with full width at
half maximum (FWHM) given by $\Delta \nu_{\mathrm{FWHM}} = \pi
C$. Using Eq.~(\ref{n_squared}), the FWHM of the $n$th harmonic
of the step recovery diode, $\Delta \nu_{\mathrm{FWHM}}^{n}$,
will thus be related to that of the first harmonic by
\begin{equation}
 \Delta \nu_{\mathrm{FWHM}}^{n} = n^{2}\Delta \nu_{\mathrm{FWHM}}.
\end{equation}
This broadening effect is illustrated in Fig.~\ref{SRD}(b). By
applying this relationship to the Fourier-limited linewidth
measurement of 244~\micro Hz with multiplication factor $n=10$,
we can extrapolate an upper limit to the comb intrinsic
linewidth of 2.44~\micro Hz.

\begin{figure}
\center
\includegraphics[scale = .5]{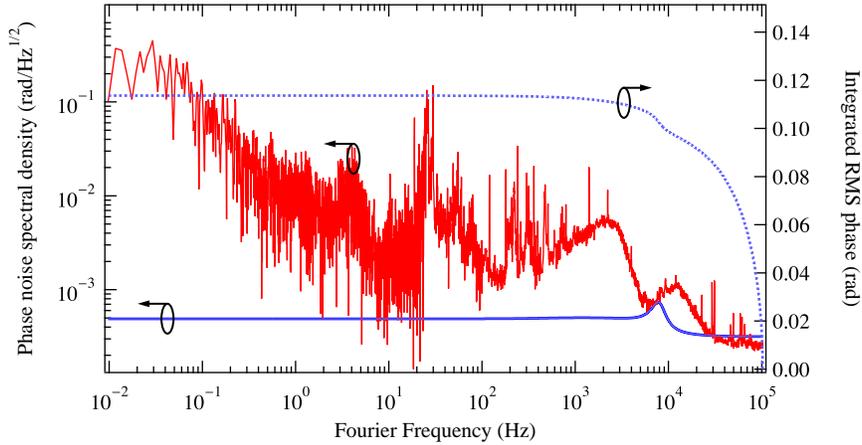}
\captionsetup{margin=0.5in,font=footnotesize, labelsep=period}
\caption{Left axis: Estimated shot noise floor (blue) compared
to the measured out-of-loop phase noise spectral density (red). Phase noise at Fourier
frequencies above 12~kHz is limited by broadband light noise
on the out-of-loop detector. Right axis: integrated root-mean-square
(RMS) phase error from 100~kHz to 10~mHz for the estimated shot
noise contribution (dashed blue).} \label{Shot_noise}
\end{figure}

Figure~\ref{Phase_noise} indicates that there is approximately
a 70~mrad difference between the extrapolated integrated phase
error and the measured integrated phase error. The extra $\sim
100$~mrad in-loop integrated phase error between 100~kHz and
10~kHz that is due only to the RF tracking filter servo bump
indicates that the true difference, which is an estimate of the
total out-of-loop noise, is at least twice as large. The extra
phase noise could come from technical sources, such as
amplitude to phase conversion in the PCF \cite{fortier2002npn}
or differential-path Doppler noise. Fundamental noise sources
that cause out-of-loop noise are shot noise and the noise term
$\epsilon_{2n}(t) - 2\epsilon_{n}(t)$.

In order to estimate the effect of shot-noise limited detection
on the out-of-loop beat, we consider the signal to noise ratio
at each of the relevant detectors assuming shot-noise-limited
detection. By modeling each servo using simple
proportional-integral (PI) transfer functions and using
empirically determined gain and bandwidth coefficients for both
the cavity length and group delay servos, we show in
Fig.~\ref{Shot_noise} the effect of shot noise on the
out-of-loop beat. This noise floor is compared with the
measured out-of-loop spectrum, indicating that below 30~kHz,
the measurement is technical noise limited. This justifies the
choice of a 100~kHz upper bound on the phase noise integration
shown in Fig.~\ref{Phase_noise}. Under a sufficiently large
servo bandwidth, the noise contribution due to
Eq.~(\ref{Lock_error}) would ideally be eliminated, leaving the
shot noise floor as the ultimate limit to the measurement's
sensitivity to other noise sources, such as out-of-loop
technical noise or intrinsic comb noise. As seen in
Fig.~\ref{Shot_noise}, the total estimated integrated phase due
to shot noise is of order 100~mrad.

While it is surprising that out-of-loop technical noise sources
do not dominate the measured out-of-loop spectrum, it is
important to note that many of the typical noise sources, such
as differential path effects, were designed to be common-mode
in our measurement. The results of \cite{schibli2008ofc} show
that even without such careful design, the total effect of
measurement noise does not inhibit sub-millihertz measurement
precision. Beyond common mode cancelation, additional
out-of-loop noise sources may be obscured due to other
higher-order correlations amongst these processes.

In analogy to the Schawlow-Townes limit, the effect of
spontaneous emission noise on frequency combs has been
explored~\cite{paschotta2004apb, paschotta2006apb, Kartner,
wahlstrand-2008}. With spontaneous emission as the only quantum
noise source, Paschotta~\textit{et~al.} find quantum-induced
timing jitter causes extra phase noise in the spectral wings
\cite{paschotta2006apb}. Wahlstrand~\textit{et~al.} take a more
complete approach, considering all quantum noise drivers and
empirically determined coupling coefficients to extrapolate the
quantum-limited linewidth of individual comb modes as a
function of frequency~\cite{wahlstrand-2008}. Both of these
results are predictions of the spectral width of a comb mode
compared to an outside reference, not between individual lines
of the same comb. However, they are useful conceptual tools to
understand the scale of the quantum noise. While a locked
in-loop error signal linewidth can be measured to be
arbitrarily small, it is the passive mode-locking mechanism
that keeps the relative coherence of comb modes an octave away
from being differentially affected by spontaneous emission
noise. We obtain a conservative lower-limit for the
free-running quantum-limited linewidth of our system from the
result given by Paschotta~\textit{et~al.}
\begin{equation}
\Delta \nu = \Delta \nu_{ST} \left[1+\left(2 \pi \delta \nu \tau_{p}
\right)^{2} \right]. \label{passive}
\end{equation}
Here, $\Delta \nu_{ST}$ is the result obtained by directly
applying the Schawlow-Townes limit to the comb, $\delta \nu$ is
the distance from the central frequency, and $\tau_{p}$ is the
output pulse width. When we insert the relevant parameters, we
obtain $\Delta \nu \simeq 100$\micro Hz for comb wavelengths
near 1064~nm and 532~nm. The analysis of Wahlstrand \textit{et
al.} indicates that full consideration of noise coupling
processes can result result in linewidths orders of magnitude
larger in these spectral regions.

By observing a relative linewidth between comb lines an octave
apart that is at least two orders of magnitude lower than that
predicted for a free-running quantum-limited comb, we have
shown that the mode-locking mechanism does an excellent job of
correlating quantum-driven phase noise between two comb modes
that are one octave apart. Even though only two degrees of
freedom are controlled, the passive mode-locking process leads
to a well-defined phase relationship across the visible and
near-IR spectrum.


\section{Conclusion}

By carefully controlling sources of technical noise, we have
placed a new limit on the phase coherence of an optical
frequency comb by using second harmonic generation to compare
modes $n$ and $2n$. Phase noise measurements show a total RMS
integrated optical phase error from 100~kHz to 10~mHz of
0.35~rad, and that the majority of accumulated phase error is
due to finite servo gains, not technical noise from the PCF or
intrinsic noise from the comb. We have additionally placed
limits on the fundamental phase-coherence of second harmonic
generation, an extension of the results of Stenger \textit{et
al.}~\cite{stenger2002umo}. The relative linewidth of comb
modes an octave apart is less than 2.5~\micro Hz, with no
significant phase noise features near the carrier, indicating
that the mode-locking process strongly correlates quantum noise
due to spontaneous emission across the comb. This result is at
least two orders of magnitude below the predicted individual
comb modes' quantum-limited linewidths. Thus, local phase
perturbation due to spontaneous emission at a given wavelength
is converted into a global phase perturbation, affecting all
modes equally within the mode-locking bandwidth. The robust
broadband phase coherence shown here demonstrates that there is
essentially no practical limit to comb-facilitated coherent
distribution of optical clock signals to arbitrary visible and
near-IR wavelengths.

\section*{Acknowledgements}
We thank J. Hall, S. Cundiff, J. Wahlstrand, and K. Cossel for
their insightful input and discussions; and C. Menyuk for
illuminating the effect of higher-order noise correlations on
our results and a careful reading of the manuscript. This work
is supported by DARPA, NIST, and NSF.

\end{document}